# From building blocks of proteins to drugs: A quantum chemical study on structure-property relationships of phenylalanine, tyrosine and dopa

Aravindhan Ganesan*, Narges Mohammadi and Feng Wang*

eChemistry Laboratory, Department of Chemistry and Biotechnology, School of Science, Faculty of Sciences, Engineering and Technology, Swinburne University of Technology, Hawthorn, Victoria 3122, Australia.

*Corresponding authors
Aravindhan Ganesan:  aganesan@daad-alumni.de
Feng Wang: fwang@swin.edu.au

# Abstract

Density functional theory and ab-initio methods have been employed to address the impacts of hydroxyl (OH) group substitutions on the physico-chemical properties of levodopa (or L-dopa) against the natural amino acids, phenylalanine and tyrosine. L-dopa, which is an important therapeutic drug for Parkinson's disease, shares structural homology with the amino acids, whose structures differ only by OH substitutions in their phenyl side chains. It is revealed that the backbone geometries of the aromatic molecules do not show apparent OH-dependent differences; however, their other molecular-level properties, such as molecular dipole moment, electronic properties and aromaticity, change significantly. The core binding energy spectra indicate that the atom sites that undergo modifications exhibit large energy shifts, so as to accommodate the changes in the intra-molecular chemical environment of the molecules. The binding energies of the modified C 1s sites in the molecules shift as much as 1.8 eV, whereas the electronic changes in their O 1s spectra happen in the higher energy region (*ca.* 536 eV). The valence spectra provide enhanced insights on the reactivity and chemical properties of the aromatic molecules. The impacts of OH moieties on the valence spectra are predominantly focussed in the energy band < 16 eV, where the frontier molecular orbitals display much reorganization and energy shifts from the amino acids to L-dopa. Of the three molecules, L-dopa also has the least HOMO-LUMO energy gap, which can readily explain its proactivity as a drug compound. Furthermore, the nuclear independent chemical shift calculations suggest that L-dopa also has more aromaticity features than those of the amino acids. The OH groups, therefore, play a more prominent role in shaping the physico-chemical properties of L-dopa, which significantly improve its drug potency.

# Introduction

Many compounds based on phenols exhibit strong antioxidant properties in biological systems. Understanding their structures, properties and interactions will help us to reveal information in a number of biochemical and physiological processes[1-4]. For example, the conversion of tyrosine (an amino acid) into dopamine (a neurotransmitter) is one of such significant biochemical transformations of aromatic amino acids[5-8]. Such the transformation is found to be initiated by an enzyme known as phenylalanine hydroxylase, which converts L-phenylalanine (L-phe) into L-tyrosine (L-tyr)[9, 10]. L-tyr can be converted into dopamine, which controls the signal transactions between nerve cells in the brain[11]. However, L-tyr is not directly biosynthesised into dopamine, but via an intermediate product known as levodopa or L-dopa or 3,4-dihydroxyl-L-phenylalanine[5, 6]. A non-heme iron enzyme, tyrosine hydroxylase, uses molecular oxygen to attach an additional hydroxyl group to the phenol moiety of tyrosine, thereby synthesising L-dopa with a catechol side chain[12]. Later, a dopa decarboxylase enzyme catalyses the decarboxylation of L-dopa, thereby synthesizing dopamine as the end product. Fig. 1 presents the sequential biochemical transformation of L-phe to dopamine.

The three molecules, that is, L-dopa and its amino acid precursors, L-phe and L-tyr, share a common phenyl structure with variations with respect to the OH group on the aromatic ring. Conceptually, L-tyr can be represented as 'L-phe + para-OH' and L-dopa can be described as 'L-tyr + meta-OH' (as shown in Fig. 1). However their physico-chemical properties are significantly different. For instance, while L-phe and L-tyr are amino acids that serve as natural protein building blocks, L-dopa is a popular catecholamine neurotransmitter drug that is widely used for the treatment of Parkinson's and other neurodegenerative disorders[6, 13-15]. L-dopa exhibits a unique drug potency to cross the blood-brain barrier. As soon as L-dopa crosses the blood-brain barrier and enters the central nervous system, it is converted into dopamine by the dopa decarboxylase enzyme (Fig. 1). In addition, L-dopa is also useful in dual-targeting the monoamine oxidase enzyme[16] and the adenosine receptors at the same time[15]. As a result, L-dopa remains an attractive drug in the fight against Parkinson's disease. A variety of L-dopa formulations are also available in the clinical market for the treatment[‡].

---

[‡] http://www.epda.eu.com/en/parkinsons/medinfo/levodopa/

L-phe and L-tyr are known to exist in a great number of low energy stable conformations, as shown by a number of experimental[17-31] and theoretical studies[32-38]. For example, Prince and co-workers recently studied the different conformers of aromatic amino acids and their effects on the electronic structures of the molecules using synchrotron sourced soft XPS techniques[17, 26]. However, it was recently shown by a combined laser desorption supersonic jet laser spectroscopy and quantum chemical studies[5, 39] that L-dopa is dominated by only a single stable conformation in the gas phase[5, 39]. The reason behind such a dramatic conformational reduction in L-dopa is not clear and the intra-molecular H-bonds are suspected as a dominating factor[39].

What is the role that the OH groups play in L-dopa, which significantly improve its drug potency? Structure-related changes in L-dopa unravel the innate physico-chemical properties of the molecule, which can be useful for understanding biological phenomena[40, 41]. Therefore, comparatively studying the structural and electronic properties of those aromatic molecules, i.e., L-phe and L-tyr against L-dopa, reveal information, which helps to understand the structure-property relationships of L-dopa.

The electronic properties of L-phe and L-tyr have been reported by different methods such as, photoemission spectroscopy (PES)[26-29, 42], near edge x-ray absorption fine structure (NEXAFS) spectroscopy[43, 44], electron energy loss spectroscopy (EELS)[45], etc. For example, Zhang et al[17] studied the electronic structures of the aromatic amino acids in the gas phase using the X-ray photoemission spectroscopy (XPS) and NEXAFS technique. The study[17] reported that the N 1s core level spectra of L-phe and L-tyr in gas phase can indicate conformer population, while their N K-edge spectra do not show conformational effects. The valence PES spectra of the aromatic amino acids have also been reported previously[26-29, 42, 46]. However, the energy and the assignment of the highest occupied molecular orbital (HOMO) of the amino acids, particularly L-phe, have often been debated by previous studies[26-29, 47, 48].

Retrieving complete information from the measured valence spectra, on the other hand, is sometimes difficult, as the measured spectral peaks are mostly broad and congested due to overlapping ionization bands. As a result, theoretical calculations can be useful to study the entire valence region of bio-molecules by relating the spectral features with their electronic structures[36, 38, 49-54]. For instance, our previous theoretical studies[36, 37] on the intra-molecular interactions in L-phe in the gas phase revealed that the inner shell of the amino acid is dominated by the functional group substitutions, while the fragment interactions are vital in its' valence electronic space. Nevertheless, there have been limited studies on the electronic structures of L-dopa, except for a recent study of L-dopa in solid phase[55].

In the present study, we analyse the effects of the hydroxyl groups on the structural, electronic and aromaticity properties of the model molecules, from L-phe and L-tyr amino acids to L-dopa, using *ab-initio* and density functional theory (DFT) methods. The aim of this work is to understand the profound changes in the properties of the molecules caused by addition of the OH groups.

**Computational details**

In the present study, the structures of the molecules are based on the conformations of L-phe[35], L-tyr[34], and L-dopa[5] reported in the literature. The conformers are re-optimised using the B3LYP/6-311G** model, which has provided accurate geometries for L-phe and other amino acids in our previous studies[37, 52, 53, 56]. The optimized structures and their nomenclatures are given in Fig. 2.

Single point calculations at the optimized structures are carried out using different quantum chemical models, such as LB94/et-pVQZ[57] for core shell and SAOP/et-pVQZ[58, 59] and OVGF/TZVP[60] for valence space. Here, the et-pVQZ basis set denotes the even-tempered polarized valence quadruple-zeta Slater-type basis set and the OVGF/TZVP is the outer valence Greens function combined with the triple zeta valence polarized basis set. The DFT based SAOP model and the Green's function based OVGF model have been efficient in calculating the accurate valence IPs for amino acids[36, 52, 56] and other bio-molecules[49, 54, 61, 62]. The former model is able to calculate the IPs in the entire valence space, while the latter is applicable for calculations of the outer valence IPs with required accuracy. Moreover, it is ascertained that the SAOP model slightly over-estimates the outermost valence IPs, especially the IP of the HOMOs, whereas the OVGF produces more accurate IPs in this region[51, 53, 56, 61]. Hence combining both the SAOP and OVGF models can, therefore, achieve good correlation between the calculated valence IPs and the experiments.

Aromaticity plays an important role in the chemical reactions and stabilities[63-

$^{65}$ of aromatic molecules. As a result, the nucleus independent chemical shift (NICS) index$^{66-68}$ is employed to study the aromaticity properties of the molecules. In this approach, a dummy atom is usually placed at the ring center (the approach is known as 'NICS(0)') or 1 Å above the ring (i.e., NICS(1)) and the magnetic shielding is calculated. The negative magnetic shielding is the NICS value. Noorizadeh and Dardab$^{69}$ have introduced the NICS-rate index that is based on the disparity of NICS indices at varying distances from the ring. This NICS-rate index is used in this study to evaluate the aromaticity features of L-phe, L-tyr and L-dopa.

All the calculations based on B3LYP/6-311G** and OVGF/TZVP models are carried out using the Gaussin09$^{70}$ computational chemistry program, whereas the LB94/et-pVQZ and SAOP/et-pVQZ calculations are accomplished using the Amsterdam density functional (ADF)$^{71}$ package.

**Results and discussion**

**A. Molecular properties**

The selected molecular properties of L-phe, L-tyr and L-dopa calculated by the B3LYP/6-311G** model are given in Table 1, while their complete geometric parameters together with available measurements$^{31, 72}$ are provided in the supplementary information SI.1). As seen in SI.1 and Table 1, no apparent changes are observed in the geometric parameters across the three molecules, including their bond lengths and bond angles. For example, the ring perimeters, $R_6$, are almost the same: 8.37 Å for L-phe, 8.38 Å for both L-tyr and L-dopa. Even the dihedral angles do not change significantly. For example, the largest change in ∠$C_{(1)}$-$C_{(\alpha)}$-$C_{(\beta)}$-$C_{(\gamma)}$, is approximately 1° from L-phe (-73.89°) to L-dopa (-72.36°).

Although the molecules do not exhibit significant geometric changes, their electronic charges have been apparently redistributed, leading to clear variations in their dipole moments. The dipole moment of L-phe is 4.88 Debye, but 3.73 Debye for L-tyr and even smaller (2.86 Debye) for L-dopa. The para-OH in L-tyr, and para-OH and meta-OH groups in L-dopa balance the charge distributions through the aromatic phenyl ring, as also indicated by the Hirshfeld charges, which are given in Fig. 2.

The Hirshfeld charges ($Q^H$) in Fig. 2 indicate very different charge distribution of the molecules. The molecules can be separated by the $C_{(\alpha)}$-$C_{(\beta)}$-$C_{(\gamma)}$ bridge, which connects the methyl carbon ($C_{(\beta)}$) of the amino acid moiety and the aromatic phenyl moieties of the molecules. The aromatic phenyl moieties of L-tyr and L-dopa serve as a buffer to accommodate the changes caused by the hydroxyl groups in L-tyr and L-dopa. For example, in L-phe, apart from the $Q^H$ on $C_{(1)}$, which is one of the six carbons in the phenyl ring and connecting directly with $C_{(\beta)}$, the charges on the other carbon atoms in the phenyl ring are more or less balanced between -0.040e and -0.051e. The $Q^H$ of $C_{(1)}$ in L-phe is given by +0.008e. In L-tyr, which has one –OH group at the para-position of the phenyl, all of the carbon atoms in the phenyl ring are affected; but the most significant changes in $Q^H$ concentrate on the $C_{(1)}$ and the para carbon, $C_{(4)}$, in Fig. 2. The $Q^H$ of the former ($C_{(1)}$) changes from positive (+0.008e) to negative (-0.001e), whereas the $Q^H$ of the latter ($C_{(4)}$) changes significantly from -0.045e (L-phe) to +0.078e (L-tyr) in order to balance the negative charge of $O_{(3)}$ in the –OH group. When the second –OH group is added on to the $C_{(3)}$ (meta) position of the phenyl ring to form L-dopa, the $Q^H$s of the "orth" carbon atoms, $C_{(2)}$ (from -0.040e to -0.061e) and $C_{(6)}$ (from -0.048e to -0.057e) positions change considerably, with respect to $C_{(4)}$ changes. Nevertheless, the $Q^H$ on one of the "meta" positions (i.e., $C_{(5)}$ site) remains little affected (from -0.056e to -0.055e).

**B. Inner shell changes**

Inner shell chemical shifts are very sensitive to local chemical environments of molecules and could serve as effective indicators that reflect the impact of OH substitution on the phenyl ring and intra-molecular hydrogen bonds in the model molecules[73]. Refer to Supplementary Information, SI.2 for all the intra-molecular hydrogen bond distances of the model molecules. Table 2 compares the inner shell vertical IPs of the aromatic molecules, L-phe, L-tyr and L-dopa, calculated using the LB94/et-pVQZ model against their respective available experimental IPs[17]. The LB94 model accurately calculates the C 1s IPs for both L-phe and L-tyr, where their ΔIP% values are smaller than 1%. It is a known fact that the core IP of the carbonyl C(=O) atom is underestimated, as it includes more electron correlations. For example, the C(=O) 1s energies of 2-azetodine were also overestimated by the LB94 and ΔE-KS methods when compared to the experimental measurements[51]. However the IP errors in the molecules become larger for the N 1s and O 1s IPs, but are still under 1%. For example, it is 0.49% for N 1s and 0.74% for O 1s, as shown in Table 2. In addition, the IP energy differences can be compensated again by applying global energy shifts to our simulations, as shown in our previous studies[37, 41, 49, 51].

Fig. 3 compares the calculated C 1s (a) and O 1s (b) spectra of L-tyr in this work against a previously reported XPS measurement[17]. The C 1s spectrum simulated with a FWHM of 0.47 eV is globally shifted to the higher energy side by 0.45 eV to match the observed peak at ca. 290.00 eV. The theoretical spectrum reproduces excellently the ratio of peaks, their intensities and widths in the measured spectrum. For instance, the theoretical C 1s spectrum (Fig. 3(a)) of L-tyr produces three peaks that are consistently decreasing in their intensities, when the energy increases, in agreement with the measured C 1s spectrum. Indeed the energy gap between the most intense C 1s peak and the middle peak is in remarkable agreement with the measured spectra, while the energy gap between the middle peak and $C_{(1)}$ 1s peak shows some discrepancies. The theory, however, underestimates the $C_{(1)}$ 1s IP by ca. 1.76 eV. The two peaks in the O 1s spectrum simulated with a FWHM of 0.97 eV (Fig. 3(b)) agree well with the measurement, after a global shift of 3.52 eV.

The calculated C 1s and O 1s spectra of L-phe, L-try and L-dopa are compared in Fig. 4(a) and Fig. 4(b), respectively. Please refer to Table 2 for their corresponding IPs. Each of the C 1s spectrum (FWHM of 0.40 eV) and O 1s (FWHM of 0.50 eV) spectrum displays a similar pattern in the three molecules, with the former showing three major peaks and the latter with two spectral peaks. The similarities in C 1s and O 1s spectra indicate that the molecules are related. The C 1s spectral peaks in Fig. 4(a) are labelled as A, B and C. IPs of these peaks decrease from A to C, but their intensities show an increasing trend (from A to C). The O 1s spectra in Fig 4. (b), on the other hand, are the opposite: both IPs and spectral intensities of the two major peaks (labelled as D and E) decrease from D to E.

In the C 1s spectra (Fig 4. (a)), the peak 'A' in the higher energy region (ca. 293 eV) is dominated by the contribution from the carbonyl carbon atom, $C_{(1)}(=O)$, in the model molecules, which is very similar to other aliphatic amino acids[56, 74, 75] and aromatic molecules[17, 37, 38]. Perhaps the most noticeable changes in the C 1s spectra of the molecules, due to the OH substitutions on the phenyl ring, are seen in the peak 'B' of the spectra, which is in a ratio of 1:2:3 for L-phe:L-tyr:L-dopa. This peak B in L-phe is assigned to the $C_{(\alpha)}$ atom[37, 38]. The carbon atoms, $C_{(4)}$ in L-tyr and $C_{(4)}$ and $C_{(3)}$ in L-dopa, which directly connect to the OH groups in the phenyl ring experience significant IP increase. For instance, the $C_{(4)}$ atom in L-tyr is attached to the $O_{(3)}$-H group that in turn increases its IP by 1.75 eV, with respect to the IP of $C_{(4)}$ in L-phe, thereby shifting the IP position of $C_{(4)}$ into the peak 'B' with $C_{(\alpha)}$ in L-tyr. Similarly, the $C_{(3)}$ and $C_{(4)}$ atoms in L-dopa are attached to the $O_{(4)}$-H and $O_{(3)}$-H groups, respectively, and hence their IPs move into the 'B' peak, along with the $C_{(\alpha)}$ atom, in L-dopa. As a result, the intensities of peak 'B' in L-tyr and L-dopa are twice or three times more than that of peak 'B' in L-phe. Such changes in the peak B of the C 1s spectra are clearly reflected in the peak 'C' as well.

The most intense peak 'C' in the lower energy side (ca. 289.5 eV) of the C 1s spectrum of L-phe denotes the strong asymmetric phenyl characters upon reduction of the high symmetry of benzene $(D_{6h})$[36]. A small extended shoulder in the 'C' peak (ca. 290 eV) is due to the bridging atoms, $C_{(\beta)}$ and $C_{(\gamma)}$, that connect the amino acid moiety with the phenyl ring. However, re-ordering of the IPs of $C_{(4)}$ in L-tyr and $C_{(4)}$ and $C_{(3)}$ in L-dopa reduce the intensities of their 'C' peaks. Therefore, the growth of the middle peak 'B' and shrink of the peak 'C' in the C 1s spectra of the aromatic molecules reflect the direct impacts of OH group substitutions in the phenyl ring. The intensity ratio of peaks, A:B:C, is given by 1:1:7 in L-phe, 1:2:6 in L-tyr and 1:3:5 in L-dopa.

The O 1s spectra of the model molecules given in Fig. 4(b), on the other hand, present two peaks (labelled as D and E) that are clearly separated by energy gaps. The peak 'E' in the lower energy side (ca. IP=534.5 eV) of the spectra is dominated by the $O_{(1)}(=C)$ atom in the molecules. This O 1s peak (i.e., peak 'E'), which corresponds to the single $C_{(1)}$ 1s peak (i.e., peak 'A' in Fig. 4(a)), is in common for the molecule series in this study. This indicates that the C=O bonding is so strong that the IPs of the carbonyl C and O atoms are distinct from the other C and O atoms, which possess single bonds in the molecules. However, it is the peak 'D' at ca. 536 eV in the O 1s spectra, which differentiates the three molecules and serves as their signatures. For example, in L-phe, this peak 'D', which is assigned to the $O_{(2)}$ atom in its hydroxyl group, exhibits the same intensity envelop with peak 'E'. But in L-tyr, this peak 'D' also receives contribution from the $O_{(3)}$ atom in the OH group in the phenyl ring. Hence the intensity of this peak 'D' of L-tyr is twice as large as peak 'D' in L-phe. Both oxygen atoms in the OH groups of L-tyr are very close in their IP energies, although one OH group is on the amino acid moiety and the other is in the phenyl ring. Such an accident "degeneracy" in IPs of the O 1s sites can be caused by their hydrogen bond networks in the vicinity of the O atoms in L-tyr (see Table SI.2).

The higher IP peak 'D' in the O 1s spectrum of L-dopa is a broad and intense peak, which differentiates L-dopa from the amino acids, such as L-phe and L-tyr. The peak 'D' of L-dopa includes contributions from the three hydroxyl oxygen atoms such as $O_{(2)}$, $O_{(3)}$ and $O_{(4)}$. The OH groups in L-dopa are in very different chemical environments so that the peak 'D' splits. The calculated energy splitting (ΔIP) is given by 0.17 eV (between $O_{(4)}$ and $O_{(2)}$) and 0.26 eV (between $O_{(2)}$ and $O_{(3)}$). This O 1s peak 'D' may split into two or even three peaks with sufficient experimental resolution. This indicates that addition of an hydroxyl group at $C_{(3)}$ (meta) of the phenyl ring makes significant impact on the electronic structure of L-dopa. It also breaks the accidental IP degeneracy in the O 1s spectrum of L-tyr. As a result, L-dopa will behave very differently from L-phe and L-tyr, supporting the fact that L-dopa is not an amino acid.

Inner shell chemical shifts indicate that the C 1s and the O 1s spectra, and therefore the structures of the molecules, are significantly affected by the OH group substitutions. Indeed the effects are more prominent on the electronic structures and molecular properties than their geometries. Nevertheless, the N 1s IP of L-phe, L-tyr and L-dopa remains unaffected with an energy value of 403.7 eV. The local chemical environment of nitrogen atom in the three aromatic molecules, L-phe, L-tyr and L-dopa, does not change. The aromatic phenyl ring serves as a buffer to accommodate the structural modifications.

**C. Effects in valence space**

Valence space information directly links to a number of chemical properties and reactivity of the molecules. Investigation of the valence spectra of L-phe, L-tyr and L-dopa is important to unveil the effects of the OH substitutions with respect to their chemical properties. Fig. 5 compares the recently measured high resolution valence XPS[26] (middle panel) of L-tyr with the theoretical spectra simulated using the SAOP/et-pVQZ (bottom panel) and the OVGF/6-311G** (top panel) models in this study. The spectroscopic pole strengths of the outer valence IPs calculated by OVGF model are ≥ 0.85 eV, demonstrating that the single particle approximation used in this model holds appropriate in the study. The theoretical spectra obtained with a FWHM of 0.50 eV are globally shifted by 1.40 eV and 0.38 eV in SAOP and OVGF, respectively; thereby aligning the HOMO peaks in the experimental and theoretical spectra at 8.47 eV. The calculated spectrum exhibits excellent agreement with the measured spectral features, such as their shapes and intensities. In fact both OVGF and SAOP models reproduce the outer valence measurements accurately, while the latter model (SAOP) continues to produce the inner valence region (i.e., > 20 eV), as shown in Fig. 5.

Fig. 6 compares the simulated valence XPS of the aromatic molecules, obtained using the SAOP/et-pVQZ model with a FWHM of 0.40 eV. The corresponding valence vertical IPs are given in supplementary information (SI.3). Although the valence spectra are more complex than the core spectra discussed before, the valence spectra of the molecules clearly display similarities and differences which can be understood from their structures. The valence spectra of the molecules again suggest that the molecules are related. Four regions of the valence spectra will be discussed in this section: innermost valence region of IP>26 eV, inner valence region of 26 eV > IPs > 18, the valence region of 18 eV < IP < 12 eV and the outer valence region of IP < 12 eV.

Three orbitals in the innermost valence region (IP>26 eV) of L-phe are dominated by the 2s electron contributions of the oxygen atoms in the carboxyl group and orbital 15a is dominated by the amino group[36, 38]. The IP energy gaps between 13a and 14a (from oxygen atoms in the carboxyl) and 14a and 15a are 2.06 eV and 2.67 eV, respectively (see SI.3 and also Fig. 6). The innermost valence orbital of L-tyr (14a) and L-dopa (15a) are also dominated by the carboxyl group, as shown in Fig. 6, while the IP peaks of orbital 17a (in L-tyr) and orbital 19a (in L-dopa) are dominated by the amino contributions. Although more new peaks appear in this valence region (particularly between 30-32 eV) in L-tyr and in L-dopa, due to additional OH groups, they do not noticeably change the original three peaks which are related to 13a, 14a and 15a of L-phe. The additional new peaks, at 32.26 eV in L-tyr and at 30.86 eV and 31.58 eV in L-dopa, are highlighted and their orbital diagrams, which belong to the oxygen atom(s) of the hydroxyl groups on the phenyl ring of L-tyr and L-dopa, are also given in Fig. 6. The innermost valence MOs (molecular orbitals), which are dominated by the functional groups, in the three molecules are, therefore, almost located in the same energy position in the valence spectra (refer to Fig. 6). This again indicates that changes in the phenyl ring exhibit minimum effects on the carboxyl groups and amino group of the molecules, which has also been illustrated by the core

IP spectra presented in the previous section.

The spectral peaks in the region of 26 eV > IPs > 18 of the three molecules align excellently. The orbitals in this region are mostly populated by the C 2s electrons, however, with the exception of the peaks at ca. 18 eV – 20 eV of L-tyr and L-dopa that also include small p electron contributions from their OH groups on the phenyl ring. The region of 18 eV > IPs > 12 eV, which includes orbitals 24a-40a in L-phe, orbitals 26a-45a in L-tyr and orbitals 28a-48a in L-dopa, show significant delocalized molecule-dependent bonding. This chemical bonding region has been observed previously in nucleosides[49] and other amino acids[56]. More information regarding the orbital densities on the region (see SI.4 –SI.6) indicates that strong 2p electron bonding interactions are from the side chain of the molecules. Therefore, this mid-valence region (i.e., 18 eV > IPs > 12 eV) that displays complex delocalized picture serve as the fingerprint valence region of the molecules.

Frontier molecular orbitals (MOs) in the outer valence region of the spectra provide useful information to understand the properties of the molecules. These outer valence MOs contribute two peaks 41a and 42a-44a for L-phe, three peaks 45a, 46a-47a and 48a for L-tyr and three peaks again, 49a, 50a-51a and 52a for L-dopa, as indicated in Fig. 6. While the peak at ca. 11.66 eV is almost unchanged in the molecules, the larger L-phe peak splits into two peaks each in L-tyr and L-dopa. Fig. 7 details the frontier orbitals of the molecules using their molecular energy diagram. As seen in this figure, the outermost valence region < 12 eV of the three aromatic molecules involves four MOs and the lowest unoccupied molecular orbital (LUMO). Although the LUMOs of the molecules exhibit similar bonding nature, the highest occupied molecular orbital (HOMO) of L-phe is quite different from the HOMOs of L-tyr and L-dopa. The HOMO (44a) of L-phe exhibits electron densities from all over the molecule including the phenyl ring and the amino acid moiety. When the hydroxyl group(s) replaces the para-H (and meta-H) in L-tyr and/or L-dopa, the HOMO electron density of the amino acid moiety is significantly reduced and is transferred to the phenyl ring and the hydroxyl groups as shown in Fig. 7. As a result of the transferred electron density, the IP of the HOMO reduces in L-tyr and reduced further in L-dopa, so that the HOMO-LUMO energy gap reduces from L-phe (4.81 eV) to L-tyr (4.22 eV) and to L-dopa (4.10 eV). It is noted that the HOMO-1 (47a) and HOMO-2 (46a) orbitals of L-tyr are swapped with respect to those in L-phe (43a and 42a), as shown in Fig. 7.

**D. Aromaticity**

The frontier MO analyses show noticeable changes in the 'π' characters of L-phe, L-tyr and L-dopa. These changes indicate that the aromatic characters of the three molecules may be affected as the result of OH groups in their phenyl rings. Further analyses are warranted to understand the changes in their aromaticity properties. This study employs NICS-rate index method to evaluate the aromaticity features of L-phe, L-tyr and L-dopa.

Fig. 8 compares the calculated NICS-rate curves of L-phe, L-tyr and L-dopa. The minima and the maxima of the NICS-rate curves of the three molecules correlate very well, where the minima is reached at a distance of 0.6 Å above the ring and the maxima at 1.6 Å above the ring. Indeed the NICS-rate curves display some trends. Among the three molecules, L-phe displays a large minimum and a large maximum. In the region of < 1.4 Å, the NICS-rate values of L-dopa are the highest, followed by those of L-tyr and then by L-phe, i.e., L-dopa > L-tyr > L-phe. In the region of 1.4 Å – 3.4 Å, the NICS-rate trend becomes L-dopa ≃ L-tyr < L-phe and finally at any distance > 3.4 Å above ring center, the NICS values of all the three molecules are approximately the same (i.e., L-dopa ≃ L-tyr ≃ L-phe). Higher NICS-rate values indicate larger aromaticity of the molecule at the given distance. Fig. 8 indicates that at small distances (< 1.4 Å), L-dopa is more aromatic, whereas L-phe is more aromatic at larger distances. To this regard, dimensionless parameters known as NICS-rate ratio (NRR)[69] and NRR$_{(\sigma)}$ were also calculated. NRR is calculated by taking the absolute ratio of the maximum and the minimum NICS-rates, as shown by a previous study[69]; whereas, the NRR$_{(\sigma)}$[46, 76] in this work can be described as,

$$NRR_{(\sigma)} = \left| \frac{(NICS\_rate_{(MAX)}) * r_{(MAX)}}{(NICS\_rate_{(MIN)}) * r_{(MIN)}} \right|$$

The calculated NICS(0), NRR and NRR$_{(\sigma)}$ of the molecules are presented in Table 3. The NRR$_{(\sigma)}$ values display an increasing trend from L-phe to L-dopa, i.e., 3.12 (L-phe) < 4.51 (L-tyr) < 5.99 (L-dopa), which agrees with their NICS(0) and NRR values as well (refer to Table 3 for values). All the measures together indicate that the aromaticity properties increase with the increasing numbers of OH group substituents. According to the Keto-enol tautomerism of the phenol group[77], increase in the number of hydroxyl groups in the phenyl ring leads to the extended aromatic conjugation. Our results agree very well with this keto-enol tautomerism concept, indicating that L-dopa is more aromatic than its amino acid precursors, L-tyr and L-phe. Increased aromaticity can be yet another feature shown by this study to enhance the physico-chemical properties of L-dopa as a drug.

**Conclusions**

The effects of OH substitutions on the molecular and electronic properties of the aromatic molecules, from L-phe and L-tyr amino acids to L-dopa, a drug for neurotransmitter disorders, have been investigated using quantum mechanical methods. The geometrical parameters, including the perimeter of the phenyl ring ($R_6$), are less affected by the OH substitutions, while other molecular properties, such as ionization spectra and aromaticity properties of the molecules display considerable changes. The OH substitutions affect the charge redistribution in the molecules, which is apparent from the reduction of dipole moments from L-phe to L-dopa.

The intra-molecular interactions of the aromatic molecules are revealed using the theoretical photoelectron spectroscopy. It is found that the hydroxyl groups on the phenyl ring cause some property changes locally. The phenyl aromatic ring serves as a buffer to resist the changes, while the amino acid moiety of the molecules acts almost independently from the phenyl ring. The C 1s spectra of L-tyr and L-dopa differ from the spectrum of L-phe by the intensity of the middle peak (ca. 291.50 eV), which in fact increase at the price of the intensity of the lowest energy peak (ca. 289.50 eV). The O 1s spectrum of L-dopa features a broader as well as a more intensive peak at ca. 536 eV. The Hirshfeld charges also reflect such inner shell changes, where the modified carbon sites gain strong positive charges that affect the entire phenyl ring in the molecules, most particularly in L-dopa.

The most significant impact of the OH substitutions in the molecule are demonstrated in their valence space, especially in the mid-valence region of 12 eV – 16 eV and the frontier orbitals. The mid-valence region presents many molecule dependent spectral differences, indicating that the significant molecule-specific interactions and chemical bonding concentrate in this energy region. The frontier orbitals, on the other hand, undergo much reorganization and energy shifts in response to the OH substitutions. The HOMO-LUMO gap decreases with the increasing OH groups as, L-phe > L-tyr > L-dopa, which indicate that L-dopa can be chemically much reactive than other molecules. Nevertheless, the aromaticity features of L-dopa increases substantially against the natural amino acids. Therefere, the OH substitituions play a more prominent role in shaping the physico-chemical properties of L-dopa, which significantly enhances its' drug potencies to target neurotransmitter disorders.

## Acknowledgements

FW acknowledges her Vice-Chancellor's Research Award at Swinburne University of Technology, which supports the Postgraduate scholarship of AG. NM acknowledges her Vice-Chancellor's Postgraduate Research scholarship at Swinburne University. The authors thank Adjunct Professor Kevin Prince for sharing the measured XPS data of tyrosine that helped in verifying the accuracies of our theoretical model in this study. The supercomputing time offered by the National Computational Infrastructure (NCI) is also acknowledged.

**Table 1**: Selected molecular properties of L-phe, L-tyr and L-dopa calculated using the B3LYP/6-311G** model. Please refer to supplementary information, SI.1, for more details on the geometric parameters of the molecules.

| Parameters | L-phe | L-tyr | L-dopa |
|---|---:|---:|---:|
| $R_6$ /Å | 8.37 | 8.38 | 8.38 |
| $C_{(1)}-C_{(\alpha)}-C_{(\beta)}-C_{(\gamma)}$/° | -73.89 | -73.36 | -72.36 |
| $N-C_{(\alpha)}-C_{(\beta)}-C_{(\gamma)}$/° | 52.25 | 52.52 | 53.25 |
| µ (Debye) | 4.88 | 3.73 | 2.86 |
| $<R^2>$ /a.u. | 2163.72 | 2751.02 | 3075.92 |
| Rotational Constants | | | |
| A /GHZ | 1.69 | 1.55 | 1.21 |
| B /GHZ | 0.62 | 0.45 | 0.42 |
| C /GHZ | 0.55 | 0.41 | 0.36 |

*$R_6$ is the perimeter of the phenyl ring in the aromatic molecules in this work.

**Table 2**: Inner shell vertical IPs of L-phe, L-tyr and L-dopa calculated using the LB94/et-pVQZ model along with the available experimental data (eV).

|  | L-phe | | | L-tyr | | | L-dopa |
|---|---|---|---|---|---|---|---|
|  | **This work** | **Exp[17]** | **Δ** | **This work** | **Exp[17]** | **Δ** | **This work** |
| $C_{(1)}$ | 293.05 | 294.85 | 1.80 | 293.04 | 294.80 | 1.76 | 293.13 |
| $C_{(\alpha)}$ | 291.50 | 291.90 | 0.40 | 291.47 |  |  | 291.52 |
| $C_{(\beta)}$ | 290.07 |  |  | 290.06 |  |  | 290.12 |
| $C_{(\gamma)}$ (ring) | 290.03 | 290.30 | 0.27 | 289.93 | 290.20 | 0.27 | 290.03 |
| $C_{(2)}$ (ring) | 289.55 | 290.30 | 0.75 | 289.68 | 290.20 | 0.52 | 289.75 |
| $C_{(3)}$ (ring) | 289.63 | 290.30 | 0.67 | 289.77 | 290.20 | 0.43 | 291.45 |
| $C_{(4)}$ (ring) | 289.65 | 290.30 | 0.65 | 291.40 | 291.85 | 0.45 | 291.32 |
| $C_{(5)}$ (ring) | 289.71 | 290.30 | 0.59 | 289.67 | 290.20 | 0.53 | 289.68 |
| $C_{(6)}$ (ring) | 289.73 | 290.30 | 0.57 | 289.81 | 290.20 | 0.39 | 289.51 |
| N | 403.72 | 405.70 | 1.98 | 403.70 | 405.65 | 1.95 | 403.74 |
| $O_{(1)}$ | 534.39 | 538.05 | 3.66 | 534.38 | 538.08 | 3.70 | 534.48 |
| $O_{(2)}$ | 535.88 | 539.87 | 3.99 | 535.85 |  |  | 535.95 |
| $O_{(3)}$(ring) |  |  |  | 535.87 | 539.27 | 3.40 | 535.69 |
| $O_{(3)}$(ring) |  |  |  |  |  |  | 536.12 |

**Table 3**: The calculated NICS(0), NRR and NRR$_{(\sigma)}$ values for L-phe, L-tyr and L-dopa.

| NICS values | L-phe | L-tyr | L-dopa |
|---|---|---|---|
| **NICS(0)** | -8.67 | -9.49 | -10.45 |
| **NICS-Rate (MAX)** | 9.94 | 8.60 | 8.71 |
| **r(MAX)** | 1.60 | 1.60 | 1.60 |
| **NICS-Rate (MIN)** | -8.49 | -5.08 | -3.87 |
| **r(MIN)** | 0.60 | 0.60 | 0.60 |
| **NRR** | 1.17 | 1.69 | 2.25 |
| **NRR$_{(\sigma)}$** | 3.12 | 4.51 | 5.99 |

**Figure captions**

**Fig. 1**: Schematic representation of the biochemical transformation of aromatic amino acids into dopamine.

**Fig. 2**: Optimized structures of L-phe, L-tyr and L-dopa, along with their nomenclatures and Hirshfeld charges calculated using LB94/et-pVQZ model in this work. Numberings 1-6 and α, β, and γ denote the carbon atoms. The colours on the atom represent the charge distribution, which can be seen in the online version of the manuscript.

**Fig. 3**: Comparison of the theoretical (LB94-et/pVQZ in this work) and experimental[17] C 1s (a) and O 1s (b) spectra of L-tyr. The theoretical spectra are simulated with an FWHM of 0.47 eV and shifted by 0.45 eV (for C 1s) and 3.52 (for O 1s) to match the experiment.

**Fig. 4**: Comparisons of the C 1s spectra (a) and O 1s spectra (b) of L-phe, L-tyr and L-dopa simulated with an FWHM of 0.4 eV and using the LB94/et-pVQZ model.

**Fig. 5**: Comparison of the valence spectra of L-tyr simulated in this work (FWHM=0.5 eV) using the SAOP/et-pVQZ and OVGF/6-311G** against a previous experimental spectrum[26].

**Fig. 6**: Valence ionization spectra (FWHM=0.5 eV) of L-phe, L-tyr and L-dopa based on the SAOP/et-pVQZ calculations in this work. The MOs of the corresponding peaks are marked in the spectra and a few orbital diagrams in the inner valence region are also shown. Please refer to supplementary information, SI.3, for the numerical values.

**Fig. 7**: Frontier orbital correlation diagrams of L-phe, L-tyr and L-dopa along with the HOMO-LUMO energy gaps of the molecules.

**Fig. 8**: NICS-rate graph of L-phe, L-tyr and L-dopa calculated as a function of distance (Å).

**Fig. 1:**

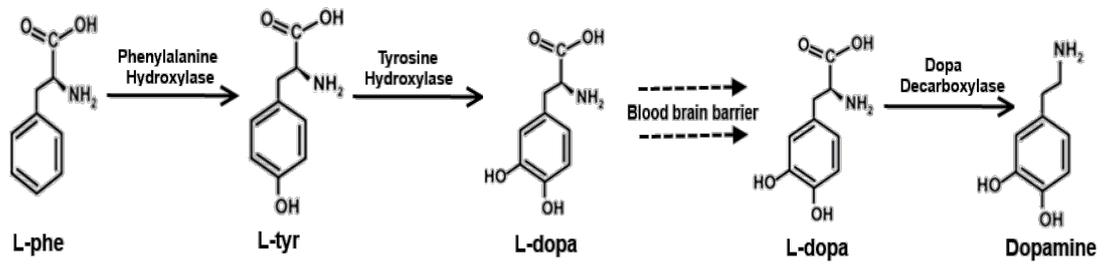

**Fig. 2:**

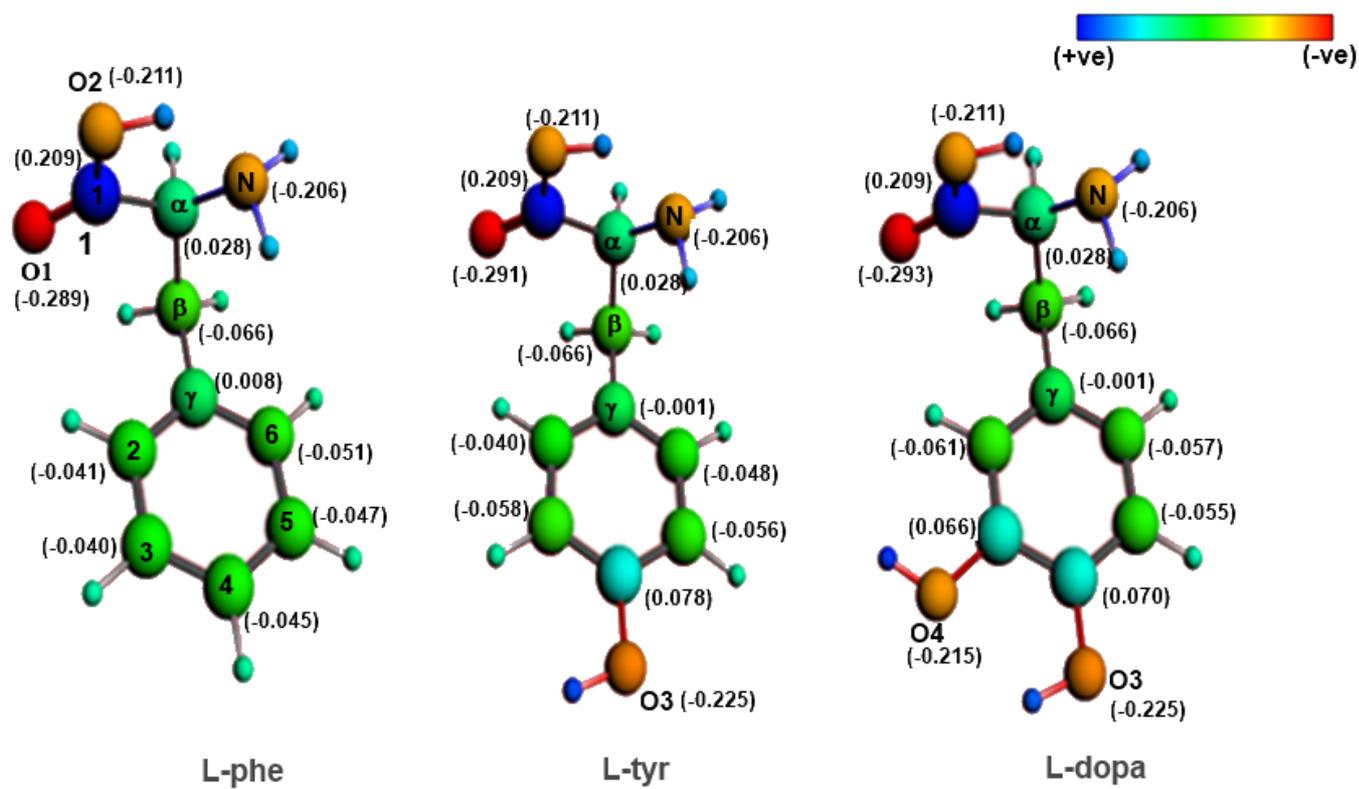

**Fig. 3:**

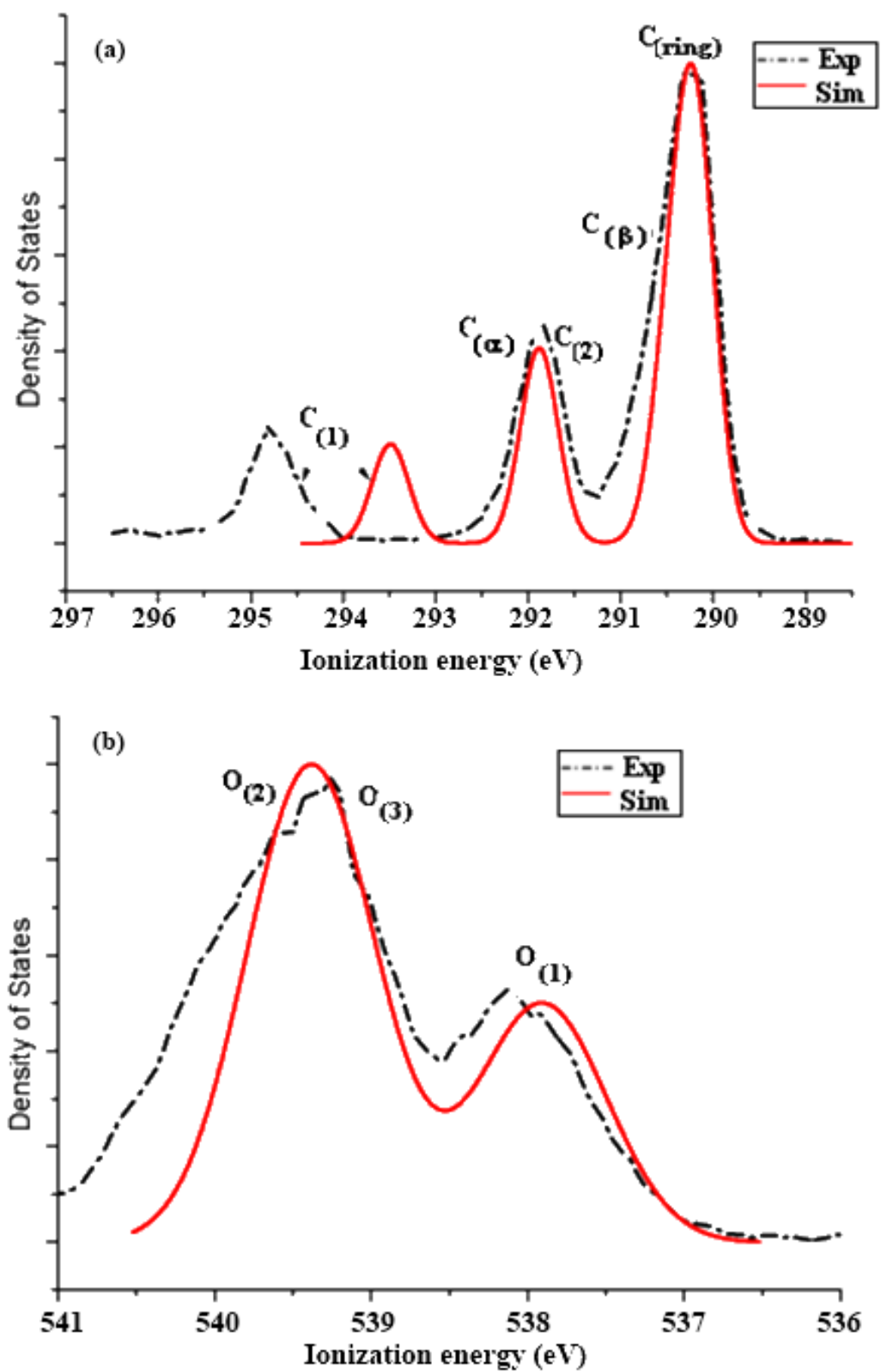

**Fig. 4:**

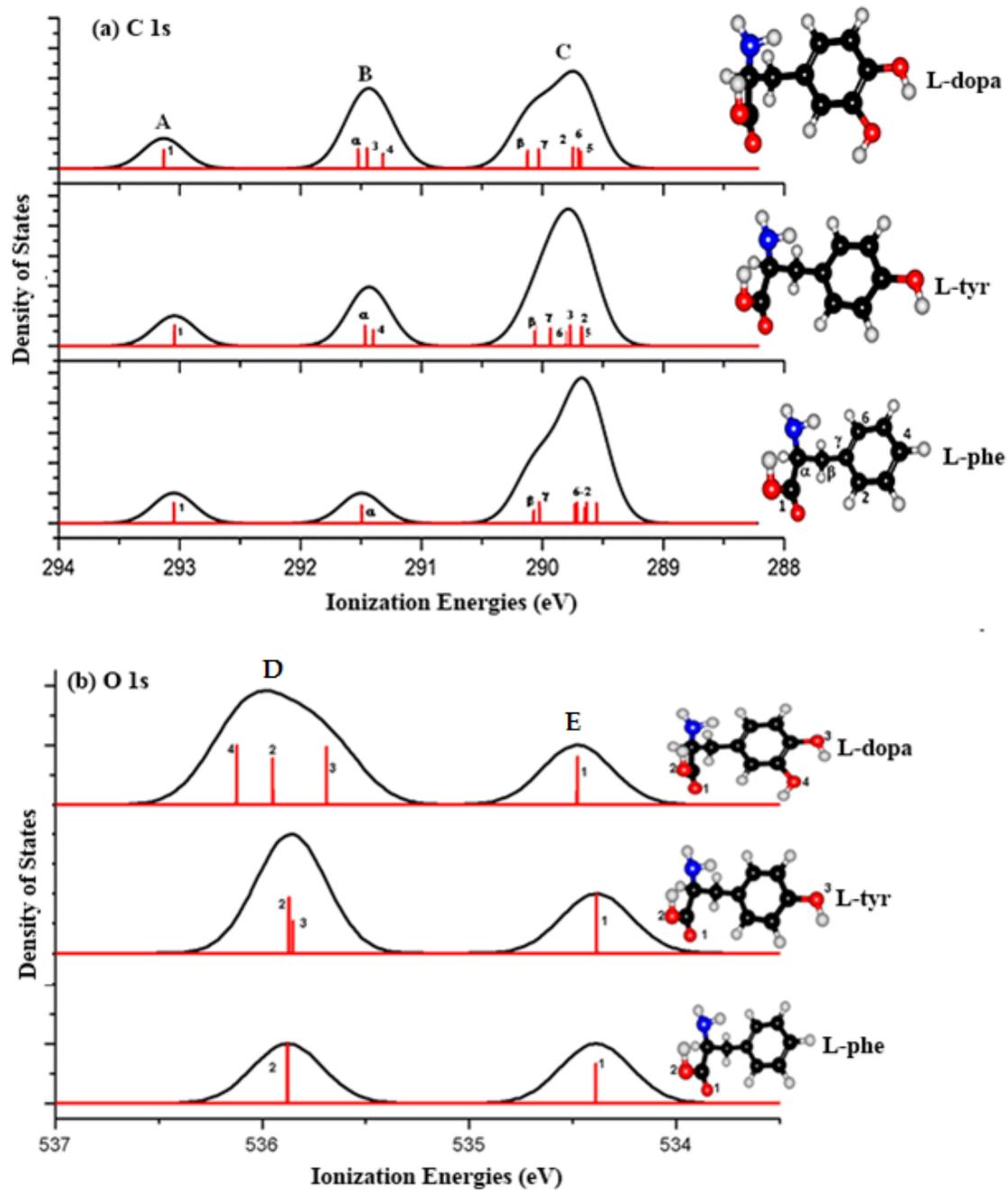
23

**Fig. 5:**

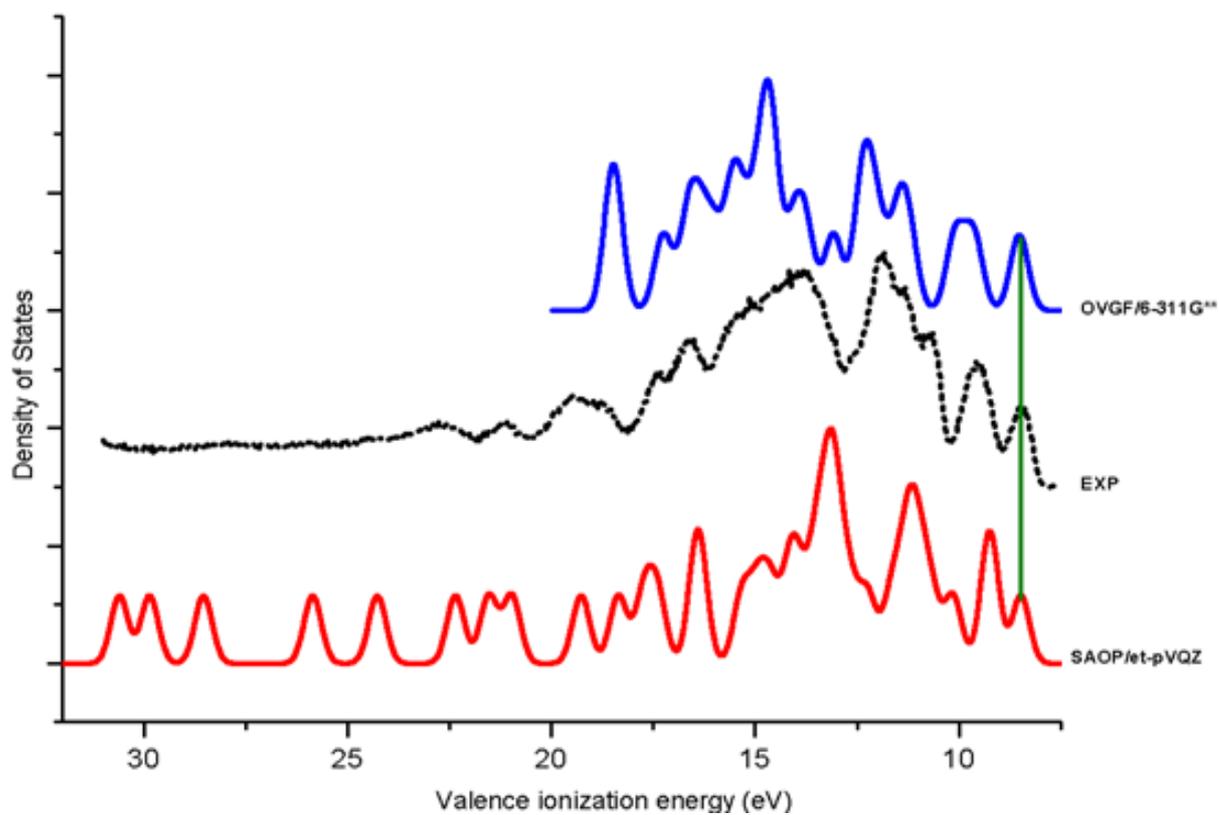

**Fig. 6:**

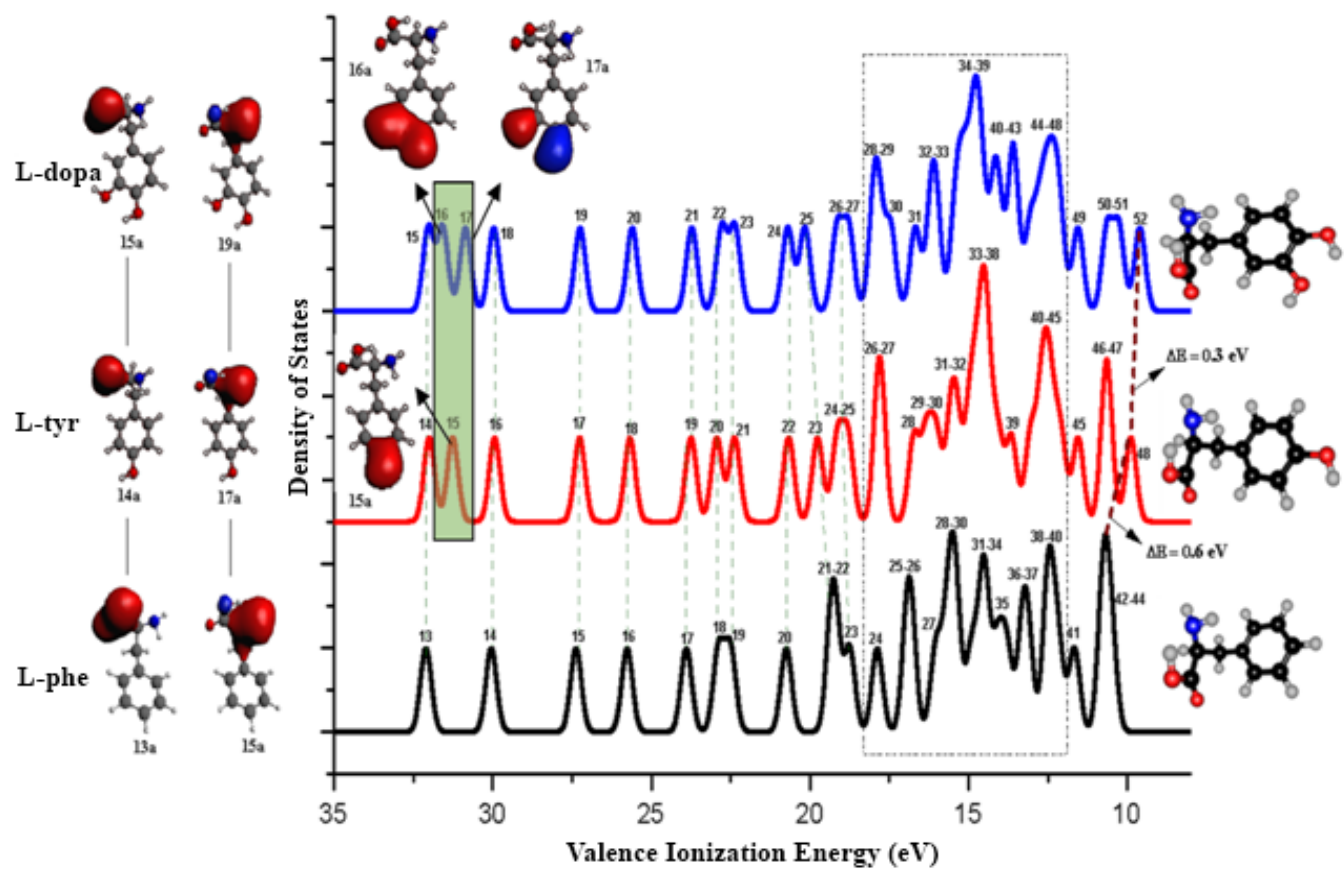

**Fig. 7:**

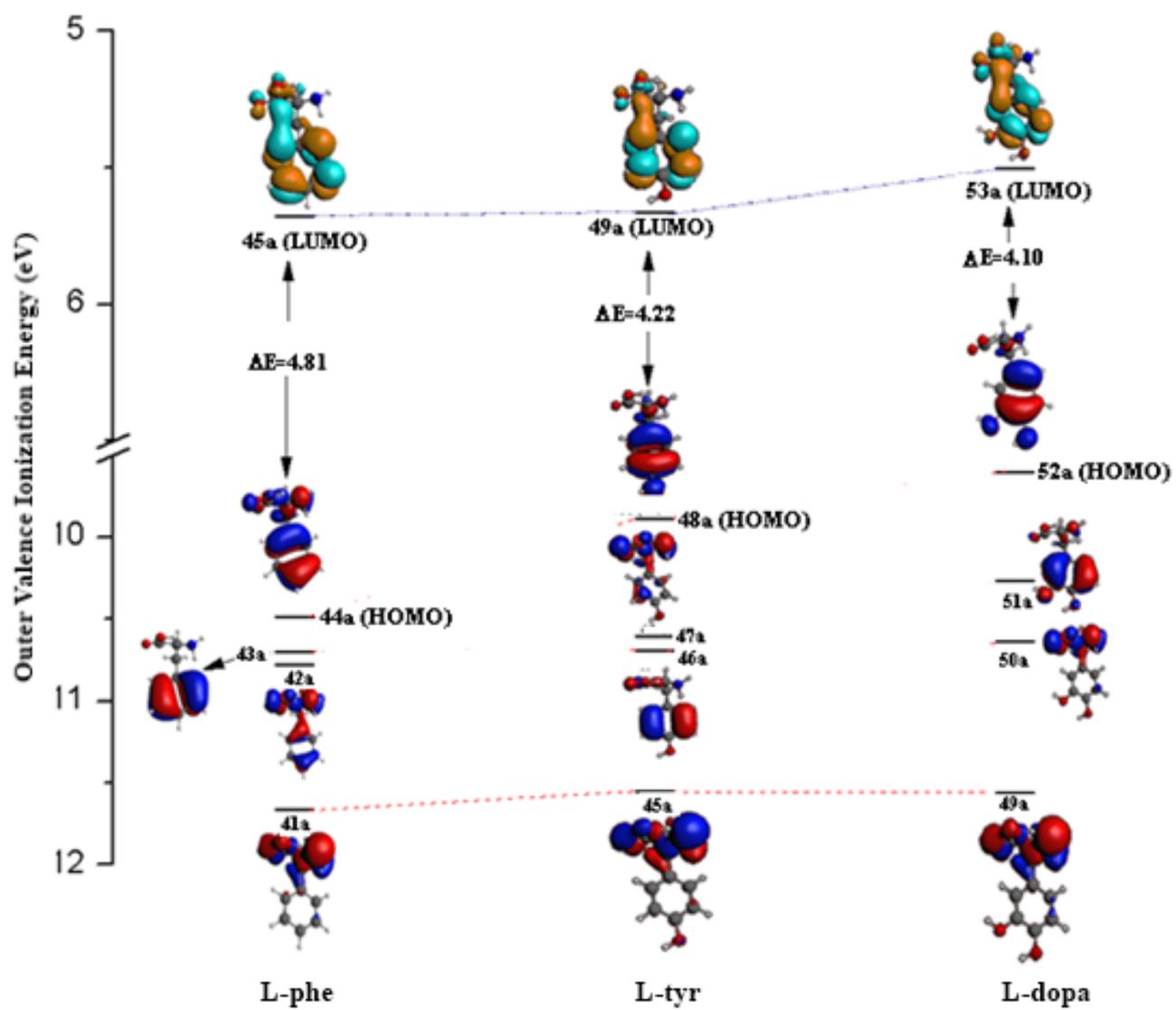

**Fig. 8:**

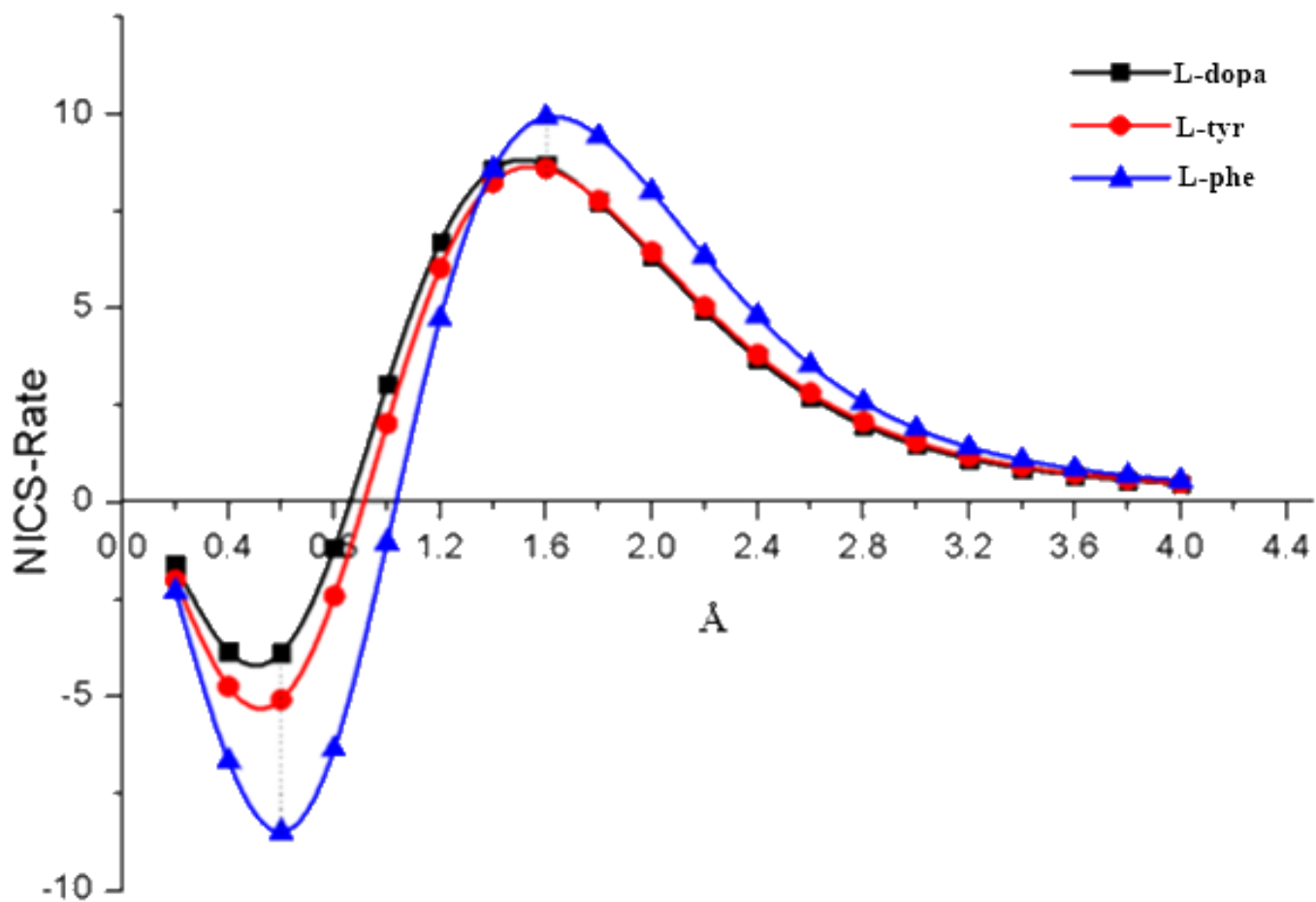